# A theoretical study of the structural phases of Group 5B – 6B metals and their transport properties

N. O. Nnolim, T. A. Tyson

*Department of Physics, New Jersey Institute of Technology,*

*Newark, New Jersey 07102*

L. Axe

*Department of Civil and Environmental Engineering, New Jersey Institute of Technology,*

*Newark, New Jersey 07102*

**ABSTRACT**

In order to predict the stable and metastable phases of the bcc metals in the block of the Periodic Table defined by groups 5B to 6B and periods 4 to 6, as well as the structure dependence of their transport properties, we have performed full potential computations of the total energies per unit cell as a function of the $c/a$ ratio at constant experimental volume. In all cases, a metastable body centered tetragonal (bct) phase was predicted from the calculations. The total energy differences between the calculated stable and metastable phases ranged from 0.09 eV/cell (vanadium) to 0.39 eV/cell (tungsten). The trends in resistivity as a function of structure and atomic number are discussed in terms of a model of electron transport in metals. Theoretical calculations of the electrical resistivity and other transport properties show that bct phases derived from group 5B elements are more conductive than the corresponding bcc phases, while bct phases formed from group 6B elements are less conductive than the corresponding bcc phases. Special attention is paid to the phases of tantalum where we show that the frequently observed β phase is not a simple tetragonal distortion of bcc tantalum.

# I. INTRODUCTION

Under equilibrium conditions free standing metals in groups 5-6 of the Periodic Table form body centered cubic (bcc) structures. The bcc structure is a special case of the general body centered tetragonal (bct) structure in metals. During the preparation of thin films of bcc metals on various substrates, conditions such as impurity levels in the preparation environment, substrate strain and substrate temperature can result in structural phases other than bcc being formed. Accordingly, numerous investigators have prepared, mainly by vacuum evaporation and deposition techniques, thin film bct phases of cubic metals, which were found to be metastable (*i.e.* which reverted back to the bcc phase upon heating or further deposition). These phases usually have mechanical and transport properties different from those of the corresponding bcc phase. The exact structural forms of these phases are usually not immediately deducible from standard experimental analytical techniques alone. An insightful analysis usually requires applying theoretical methods in the form of *ab initio* band structure calculations in addition to typical experimental analytical tools. In order to understand and predict the bct structures formed by thin films of bcc transition metals in columns 5–6 of the Periodic Table, we have carried out total energy calculations as a function of the $c/a$ ratio for these metals, while keeping the unit cell volumes constant at the experimental values. We have also computed the electrical resistivity and the plasma frequency, important quantities in the areas of transport and superconductivity. The electrical resistivity is a readily obtainable characteristic property and the plasma frequency is directly proportional to the zero temperature conductivity of metals. The temperature dependence of the electrical resistivity is calculated for all the studied metals except Cr, for reasons which are described in part A of the theoretical background section of this paper. The zero temperature plasma frequency is calculated for all the studied bcc metals. Its pressure dependence is calculated for Nb and Ta. Our results are compared to other calculated values as well as values obtained from experimental measurements at low temperatures, and in the case of the resistivity, our results are compared to experimental data.

Amongst the metals we have studied, metastable phases in elemental form of tantalum,[7–9,18,23-31] tungsten[4,6] and vanadium[1,2] have been reported in the literature. These metastable phases are usually formed most easily as thin films and have been observed to contain a significant amount of defects and non-crystallinity. They also exhibit a number of useful electrical phenomena, for example, the superconducting transition temperatures in metastable phases of tungsten and tantalum are believed to increase appreciably relative to the bulk bcc values.[45,47] A brief discussion of the metastable phases of the metals in this study, which have been reported in the literature follows.

**A. Metastable phase of tungsten**

Tungsten films are technologically important because they are used as x-ray masks, metallization layers and interconnects on silicon-based devices. Several authors[4,6,20] have reported a metastable form of tungsten with a lattice constant of approximately 0.504 nm, which crystallizes in the A-15 structure and is referred to as β tungsten. The A-15 β-W phase has been prepared by sputter deposition,[4,48,49] evaporation,[47] and reduction of tungsten oxides.[50,51] This metastable phase has been reported to transform to the stable cubic A-2 α-W upon heating, although the reported transition temperature ranges vary widely. The oxygen content of the A-15 β-W phase is believed to influence the transition temperature range, with higher oxygen content films requiring longer transition times and higher heating temperatures. The A-2 α-W structure formed from A-15 β-W in this manner has the same oxygen content and XPS binding energy as its precursor, but much lower resistivity, on the order of 80% less.[6,20,21] The decrease in resistivity has been ascribed to the elimination of the β-W defect structure and an increase in particle size. Another work[5] has reported that the A-15 β-W phase systematically evolved to the A-2 α-W phase with decreasing oxygen concentration and lattice parameter. The stability of the A-15 β-W was also determined to decrease with decreasing oxygen concentration, implying that oxygen impurities stabilize the A-15 β-W phase.

### B. Metastable phase of vanadium

Tian and coworkers[1,2] have used the FP LAPW method in the DFT formalism to carry out total energy calculations for vanadium as a function of the $c/a$ and $V/V_0$ ratios, where $V_0$ is the equilibrium volume. They predicted a metastable phase of vanadium with $c/a$ and $V/V_0$ ratios of 1.78 and 1.02 respectively, and then proceeded to prepare metastable phases of vanadium on different substrates, which they identified as strained states of the theoretically predicted metastable phase. In the first of those experiments,[1] vanadium was deposited on Ni{001} with an $a$ lattice parameter 0.8% smaller than the theoretically predicted square side (2.51Å) of tetragonal vanadium. The vanadium films could be grown only up to thicknesses of about 20 Å before defects were observed. Since the lattice parameter of Ni{001} is smaller than that of the predicted tetragonal phase of vanadium, the vanadium film was believed by the authors to be compressively stressed. Making use of the Bain path[18] concept, the authors conjectured that using a substrate with a lattice parameter slightly larger (by 2%) than that of tetragonal vanadium would result in more ordered vanadium thin films undergoing tensile stress, and chose Cu{001}($a$ =2.56 Å) as a substrate. Quantitative Low Energy Electron Diffraction (QLEED) analysis was subsequently used to determine that very thin films of tetragonal vanadium were formed on the Cu{001} substrate, with structures derived from strain on the theoretically predicted metastable tetragonal phase of vanadium. The films had many defects and poor crystallinity and eventually collapsed into {110}bcc vanadium. The high density of disorder and defects in the thin film of tetragonal vanadium was attributed to the small energy barrier between the strained metastable state and the bcc ground state of vanadium. To our knowledge no similar study has been carried out for tantalum.

### C. Metastable phase of tantalum

The formation of tantalum metastable phases on a variety of substrates has been reported in a large number of sputtering experiments.[7,9,18,23-31] The metastable phase, referred to as the β phase, is sometimes formed alongside the stable bcc phase, referred to as the α phase. The β phase of tantalum occurs mostly as a thin film and is not easily formed as bulk material. It is harder, more brittle, and less ductile than the bcc phase. It is known to have a 45% smaller average grain size than bcc tantalum, and to frequently contain a high concentration of impurities and defects, in addition to having a resistivity (170 - 210 μΩ cm) much higher than that (15 – 60 μΩ cm) of the stable bcc phase.[23,24,30,45,46,53] The β phase of tantalum has been reported to nucleate preferentially on certain kinds of substrates,[25-29] and may be stabilized by oxygen impurities.[7,25,26,29-31]

From work such as that of Tian *et al*,[1,2] and Marcus *et al*,[12] who have used total energy calculations to show that the bulk structure of a thin Ti film on an Al{001} substrate is strained fcc Ti, it is evident that the nature of the substrate on which a thin film is deposited can determine the kind of stress prevailing in the deposited film, and therefore the final structural form of the film. β-tantalum has been observed to form preferentially on certain substrates, so the nature of the substrate upon which thin tantalum films are deposited appears to play an important role in the subsequent formation of the β phase. It is not known however whether metastable tantalum structures are formed at any stage of the production of β tantalum in a substrate-strain mediated process analogous to what has been related above for vanadium. In this paper, we try to predict the metastable tantalum structure that would result from such a process, calculate its transport properties and then compare them to the observed transport properties of β-tantalum. The primary motivation for this work then is to investigate whether a tetragonal distortion of the stable bcc structure of tantalum at constant volume forms a metastable phase that can account for the large resistivity of β–tantalum relative to α bcc tantalum. We do this by calculating, for tantalum as well as for the rest of the bcc metals in groups 5B – 6B and periods 4 - 6 of the Periodic Table, the total energy per unit cell as a function of the $c/a$ ratio, keeping the unit cell volumes constant at the experimental values. The electrical resistivities are then calculated for all the bcc metals and their predicted bct phases using the Lowest Order Variational Approximation (LOVA) formalism for the solution of the Bloch-Boltzmann transport equation. Total energy calculations as a function of the *c* tetragonal lattice constant have therefore been carried out for the metals V, Nb, Ta, Cr, Mo and W in the block of the Periodic Table defined by groups 5B to 6B, and periods 4 to 6. In all cases a metastable bct phase was predicted. Different alloy combinations of the metals in this study have been prepared as metastable phases before,[14-17] but to our knowledge, except for V, Ta and W, no metastable bct phases of these metals have been prepared in elemental form. Hence, we hope that this study will stimulate experimental work on the metastable phases of groups 5B and 6B metals.

In section II of this paper, the theoretical background material is reviewed. In section III, the computational method we have used is described in detail. In section IV, our results are presented. In section V, these results are discussed and then we conclude in section VI.

## II. Theoretical Background

We now give the theoretical background needed for the discussion of the transport and electronic properties studied in this work.

### A. Electrical Resistivity

At temperatures above $\Theta_D$, the Debye temperature, the dominant contribution to the electrical resistivity of pure metals comes from electron-phonon scattering. The electrical and thermal resistivities of metals are well described by LOVA, which is most accurate for $T \geq \Theta_D$, where $T$ is the temperature. LOVA neglects the energy dependence and anisotropy of the electronic distribution function, and therefore assumes that the Fermi surface displaces rigidly in response to an external field.[37] In the LOVA formalism the electrical resistivity is given as an upper bound to the true resistivity by the expression[37]

$$\rho(T) = \frac{6\pi V k_B T}{e^2 \hbar N(\varepsilon_F) <v^2>} \int_0^\infty \frac{\alpha^2_{tr} F(\omega) x^2}{\sinh^2 x} \frac{d\omega}{\omega}, \tag{1}$$

where $x = \hbar\omega/2k_B T$, $V$ is the volume per unit cell and $\alpha^2_{tr} F(\omega)$ is the transport spectral function given by

$$\alpha^2_{out} F(\omega) - \alpha^2_{in} F(\omega). \tag{2}$$

In the high temperature limit, the expression (1) for the resistivity becomes

$$\rho = \frac{6\pi V k_B T}{e^2 \hbar N(\varepsilon_F) <v^2>} \lambda_{tr} \left(1 - \frac{\hbar^2 \langle \omega^2 \rangle_{tr}}{12 k_B^2 T^2} + \cdots \right). \tag{3}$$

The resistivity can then be rewritten as

$$\rho = (4\pi/\omega^2)(1/\tau), \tag{4}$$

where $\omega$ is the plasma frequency and $1/\tau$ is the electron-phonon scattering rate given by

$$1/\tau = 2\pi k_B T \lambda_{tr} \left(1 - \frac{\hbar^2 \langle \omega^2 \rangle_{tr}}{12 k_B^2 T^2} + \cdots \right), \tag{5}$$

with the electron-phonon interaction parameter $\lambda_{tr}$ given by

$$\lambda_{tr} = 2\int_0^\infty \alpha^2_{tr}(\omega) F(\omega) \frac{d\omega}{\omega}. \tag{6}$$

The mass enhancement factor $\lambda$ determines the superconducting transition temperature $T_c$ according to McMillan's formula[61]

$$T_c = \frac{\Theta_D}{1.45} \exp\left(-\frac{1.04(1+\lambda)}{\lambda - \mu^*(1+0.62\lambda)}\right). \tag{7}$$

$\lambda_{tr}$ is closely related to the mass enhancement factor $\lambda$ by[52]

$$\lambda_w = \frac{3\sum_{k,k'} w^{k,k'} |M^{k,k'}|^2 v_{kx} v_{k'x} (\hbar\omega_{k-k'})^{-1} \delta(\varepsilon-\varepsilon_k)\delta(\varepsilon-\varepsilon_{k'})}{N(\varepsilon_F) <v^2> \sum_{k,k'} w^{k,k'} \delta(\varepsilon-\varepsilon_k)\delta(\varepsilon-\varepsilon_{k'})}, \tag{8}$$

where $|M^{k,k'}|$ is the electron-phonon matrix element of scattering from state $k$ of energy $\varepsilon_k$ and group velocity $v_k$ to a state $k'$ by emission of a phonon of energy $\hbar\omega_{k-k'}$. The weighting factor $w^{k,k'}$ is 1 for $\lambda$ and $(v_{kx}-v_{k'x})^2$ for $\lambda_{tr}$. In an isotropic system, $\lambda_{tr}$ and $\lambda$ differ by the factor $\langle 1-\cos\theta\rangle$ which is of the order of unity, $\theta$ being the transport process scattering angle. $\lambda_{tr}$ and $\lambda$ usually agree to within 10% for $d$ band metals,[37,52] so we approximate $\lambda_{tr}$ by $\lambda$, which is obtained from electronic tunneling data and McMillan's formula for $T_c$.[34]

We attempt to use our calculations to reproduce the experimental resistivities of all the studied bcc metals (except Cr) over a defined temperature range. We also try to reproduce the observed trend in the magnitudes of the resistivities at room temperature in order to have some degree of confidence in our ability to use the calculations to differentiate between possible structures of β-Ta based on comparison of calculated resistivity values to experimental data. The method we are using to calculate the resistivities does not permit calculations for Cr, because bcc Cr is not a superconductor. Accurate resistivity data[56] for the rest of the metals, from which the residual resistivities have been subtracted, have been used for comparisons to our calculations.

The room temperature resistivities of β-Ta thin films with thicknesses ranging from 100 – 20,000 Å have been measured by the four-point probe method[23,24,30,45,46,53] and the values obtained fall in the range 170 – 220 μΩ cm. The value of the bcc Ta room temperature resistivity is 13.6 μΩ cm,[24] 1200 – 1600 % less than the range of values that have been reported for β-Ta. It is therefore crucial that the errors associated with our calculated resistivities remain well below these boundaries.

### B. Plasma Frequency

The plasma frequency $\omega$ is defined by[33]

$$\hbar^2\omega^2 = \frac{8\pi e^2}{(3V)4\pi\varepsilon_0} N(\varepsilon_F)<v^2>. \qquad (9)$$

The quantity $N(\varepsilon_F)<v^2>$ can be represented as

$$N(\varepsilon_F)<v^2> = \sum_{nk} |\nabla_k \varepsilon_{nk}|^2 \delta(\varepsilon_{nk} - \varepsilon_F), \qquad (10)$$

where the sums on the indices $n$ and $k$ are over occupied states.

It is illustrative to transform the sum in Eq. (10) into an integral over the Fermi surface

$$\sum_{nk} |\nabla_k \varepsilon_{nk}|^2 \delta(\varepsilon_{nk} - \varepsilon_F) = \frac{V}{8\pi^3} \int_S \nabla_k \varepsilon_{nk} dS, \qquad (11)$$

showing that the plasma frequency is directly proportional to the Fermi surface average of the velocity

$$\hbar^2\omega^2 = \frac{e^2}{12\varepsilon_0 \pi^2} \int_S \nabla_k \varepsilon_{nk} dS. \qquad (12)$$

$N(\varepsilon_F)<v^2>$ was then subsequently calculated according to Eq. (10) for all the bcc metals in this study and their corresponding bct phases using the FP-LAPW method.

## C. Densities of states

The density of electron states at the Fermi level per atom and per spin is given by

$$N(\varepsilon_F) = \frac{V}{8\pi^3} \int_{S_n} \frac{dS}{|\nabla_k \varepsilon_{nk}|}, \qquad (13)$$

therefore $N(\varepsilon_F)$ is a Fermi surface average of $1/v$, where $v$ is the electronic velocity. $N(\varepsilon_F)$ is inversely proportional to the average velocity on the Fermi surface.

## D. Pressure dependence of transport properties

Since we are attempting to predict the structure dependence of transport properties from band-structure calculations, we test the accuracy of our computational method by trying to reproduce the calculated and experimentally determined[33] pressure dependence of transport properties such as the density of electron states at the Fermi level, $N(\varepsilon_F)$, and the plasma frequency $\omega$ for the metals Nb and Ta. The experimental method used to determine the pressure dependence of the plasma frequency for Nb involved making use of the measured pressure dependence of the temperature gradient of the electrical resistivity $\partial \rho / \partial T$, published results for the superconducting transition temperature $T_c$, and the compressibility. The theoretical method consisted of a self-consistent Linear Muffin Tin Orbital (LMTO) band-structure calculation. The plasma frequency $\omega$, which is essentially the product of $N(\varepsilon_F)$ and $<v^2>$, was found to be somewhat insensitive to small changes in the Fermi energy level. The density of states at the Fermi energy, $N(\varepsilon_F)$, was determined to decrease with increasing pressure, satisfying the relation

$$\frac{\partial \ln N(\varepsilon_F)}{d \ln V} = 1.2, \qquad (14)$$

while the plasma frequency $\omega$ and $<v^2>^{1/2}$ were determined to increase with increasing pressure. The calculated energy bands were observed to broaden with increased pressure, resulting in reduced electron-electron interaction as measured by the Coulomb pseudopotential. An $N(\varepsilon_F)$ value of 10.68 states/Ry.atom.spin was obtained for Nb from our calculations, which is well within the range obtained by Neve et al.[33]

## III. COMPUTATIONAL METHOD

The WIEN97 code developed by Blaha et al[10] implementing the FLAPW method in the density-functional theory (DFT) formalism was used for these calculations. Core states are treated in a fully relativistic manner while the valence states are treated semi-relativistically. The WIEN97 code solves the single particle Kohn-Sham equations for the Kohn-Sham eigenvalues, ground state density and total energy of a many electron system using the FLAPW method without making any shape approximations to the potential. The exchange-correlation part of the total energy is represented with the GGA (generalized gradient approximation), using the parameterization of Perdew-Burke-Ernzerhof. Density-functional theory (DFT) in the GGA is now supplanting the local-density approximation (LDA) for electronic structure study and prediction of ground state properties of different materials and systems. The GGA formalism addresses the non-uniformity of the electron gas using its density gradients, while the LDA formalism assumes electron homogeneity. Using DFT in the GGA formalism, the correct bcc ground state of Fe has been predicted, a feat that the LDA formalism is unable to accomplish.[5] The total energy per unit cell was calculated as a function of the *c* tetragonal lattice constant at constant experimental volume for all the studied elements at T = 0 K. All calculations were begun by defining a unit cell, which was subsequently divided into non-overlapping atomic sphere and interstitial regions. The solutions to the Kohn-Sham equations were expanded in a combined Linear Augmented Plane Wave (LAPW) basis set consisting of plane waves in the interstitial region of the unit cell and a linear combination of products of spherical harmonics and radial functions inside the atomic spheres. The muffin-tin radii chosen for the atomic spheres ranged from 2.3 bohr for V and Cr, through 2.5 bohr for Nb and Mo, to 2.7 for Ta and W. The charge densities and the potential were expanded in partial waves up to $l = 10$ inside the atomic spheres and $l = 4$ outside the spheres. The RKMAX parameter, which is the product of the muffin-tin radius and the cutoff for the plane-wave expansion, was chosen to be 10. The magnitude of the largest vector in the Fourier expansion of the charge density, represented by the GMAX parameter, was chosen to be 14 bohr$^{-1}$. The computations were carried out using over 1800 irreducible k-points in the first Brillouin zone for all the studied elements. The k-points were generated on a grid, which was used in a modified tetrahedron integration method[11] to calculate the Fermi energy. The total charge density per unit cell was converged to within $10^{-5}$ electrons/$a_0^3$ self consistently and then converged with respect to the number of k-points in the reciprocal lattice. Thirty-nine structures, each with a different $c/a$ ratio, were created for each cubic element studied. The total energy per unit cell of each of these structures was calculated to determine the energy minima in the $(E, c/a)$ plane. Two energy minima were obtained for all the cubic elements. Over 33,000 irreducible k-points were used in the calculation of the derivatives used to determine $N(\varepsilon_F)<v^2>$ and hence the plasma frequency. The derivatives were calculated without accounting for spin-polarization. The calculation of the pressure dependence of the plasma frequency was carried out while keeping the ratio of the muffin-tin sphere volume to the total volume per unit cell constant for each structure.

## IV. RESULTS

Since some of the column VB and VIB metallic elements have unpaired $d$-shell electrons, values for the total energy per unit cell of these elements were calculated using a spin-polarized formalism, which included spin-orbit coupling. The total energies calculated in this manner for the elements were practically the same (less than 0.1% difference) as the energy calculated without spin-orbit coupling, indicating a negligible influence of spin-orbit interactions on the total energy per unit cell. Two energy minima were observed in the $(E, c/a)$ plane for all the cubic metals in this study (Figs 1(a) – 1(f)), corresponding to stable bcc and metastable bct (body centered tetragonal) phases. A metastable structure with $c/a \sim 1.4$ is also observed for all the cubic metals, corresponding to the fcc phases of the metals. The $c/a$ ratios at which bct phases are formed as well as the energy differences, $\Delta E^{TOT}$, between corresponding bcc and bct phases, are listed in Table I. Within the studied groups, the magnitude of the $c/a$ ratio at which a bct phase is formed is inversely proportional to the atomic number $Z$, while $\Delta E^{TOT}$ is proportional to $Z$. In each studied period, bct phases formed from group 5B elements had higher $c/a$ ratios than bct phases formed from group 6B elements. The $\Delta E^{TOT}$ values calculated for the group 6B elements were all greater than those calculated for the group 5B elements. The $\Delta E^{TOT}$ values for all the bcc metals ranged from 0.09 eV (vanadium) to 0.39 eV (tungsten).

### A. Densities of states

The densities of states (DOS) as functions of energy, $N(\varepsilon)$, were calculated for all the studied bcc metals, including the calculated metastable bct phases of these metals. From the $N(\varepsilon)$ vs. $\varepsilon$ figures, the total and angular momentum decomposed densities of states at the Fermi energy, $N(\varepsilon_F)$ and $N_l(\varepsilon_F)$ respectively, were computed for all phases. The total and $d$ DOS figures for the bcc and bct phases are shown in Figs. 2(a) – 2(f).

#### 1. Bcc Phases

The calculated $N(\varepsilon_F)$ for all the studied bcc metals are listed in Table II. Proceeding down the periods and also to the right across the groups of the studied block of the Periodic Table, $N(\varepsilon_F)$ was seen to decrease for the metals. The highest value of $N(\varepsilon_F)$ among the bcc elements was obtained for V, and the lowest value was obtained for W. The $d$ densities of states, $N_d(\varepsilon_F)$, were responsible for the largest contribution to $N(\varepsilon_F)$ for all the studied bcc elements, and also followed the same trend as $N(\varepsilon_F)$. The cubic symmetry splits the $d$ states further into the $d\text{-}e_g$ and $d\text{-}t_{2g}$ irreducible representations. There is a single narrow peak due to the $d\text{-}e_g$ states located in the upper conduction band region of all the DOS figures (Figs. 2(a) – 2(f)), which broadens down the periods of the studied Periodic Table block (*i.e.* as the atomic number $Z$ and the lattice constant $a$ increase). The conduction $d$ bandwidth correspondingly increases down the periods. A double peak feature due to the $d$-$t_{2g}$ states is also seen near the top of the valence band. For the group 5B elements, the double peak feature extends through $\varepsilon_F$. The double peak feature does not extend through $\varepsilon_F$ for the group 6B elements, which occurs at a minimum in the DOS for these elements. The height ratio of the $d$-$e_g$ single peak feature to the $d$-$t_{2g}$ double peak feature is highest for V and Cr, and decreases down the periods. The $d$-$t_{2g}$ states are responsible for the largest contribution to $N_d(\varepsilon_F)$ for all the studied bcc metals, slightly more so for the group 5B than the group 6B metals (Table III). The $s$ and $p$ total densities of states were insignificant compared to the $d$ and have therefore not been included in the figures.

### 2. Bct phases

The calculated $N(\varepsilon_F)$ for the bct phases derived from the group 5B bcc elements were consistently lower than those computed for their corresponding bcc phases. The reverse relationship to the bcc phase $N(\varepsilon_F)$ values was seen for the $N(\varepsilon_F)$ calculated for bct phases derived from the group 6B bcc elements. The trend in the computed $N(\varepsilon_F)$ observed for the bcc phases was mirrored by the trend in calculated $N(\varepsilon_F)$ for the bct phases, except that the calculated $N(\varepsilon_F)$ for bct Nb is less than that computed for bct Ta. The calculated $N_d(\varepsilon_F)$ for the bct phases followed exactly the same trend as the calculated $N(\varepsilon_F)$. $\varepsilon_F$ was observed to increase for all the predicted bct structures relative to their corresponding bcc structures.

The total and $d$ DOS figures for the bct phases are also shown in Figs. 2(a) – 2(f). In the DOS figures for the bct phases, the $d\text{-}e_g$ and $d\text{-}t_{2g}$ states are split further into the two $d_{z^2}$ and $d_{x^2-y^2}$ irreducible representations, and the three $d_{xy}$, $d_{xz}$ and $d_{yz}$ irreducible representations respectively by the tetragonal symmetry. The contributions to the DOS from the $d\text{-}e_g$ derived states have a double peak structure and are flattened relative to the sharp $d\text{-}e_g$ peak seen in the bcc DOS figures. The contributions to the bct DOS from the $d\text{-}t_{2g}$ derived states are also flattened relative to the double $d\text{-}t_{2g}$ peak seen in the bcc DOS figures and have a more complex peak structure. The $d\text{-}e_g$ derived states are responsible for the largest contribution to $N_d(\varepsilon_F)$ for the group 5B bct phases, while the $d\text{-}t_{2g}$ derived states are responsible for the largest contribution to $N_d(\varepsilon_F)$ for the group 6B bct phases (Table III). The complex set of $d$ bands derived from the $d\text{-}e_g$ states extend from the top of the valence band through $\varepsilon_F$ to the bottom of the conduction band. There is a double peak feature due to the $d\text{-}t_{2g}$ derived states extending through the conduction band. Significant contributions to the DOS from the $d\text{-}t_{2g}$ derived states can also be seen in the lower part of the valence band region. The ratio of the height of the $d\text{-}e_g$ derived feature to the height of the $d\text{-}t_{2g}$ derived double peak feature increases down the periods, being lowest for bct V and bct Cr and highest for bct Ta and bct W. The $\varepsilon_F$ for the bct phases derived from the group 5B elements, similar to the $\varepsilon_F$ in their corresponding bcc phases, occur in a relatively high density of $d$ bands.

The $\varepsilon_F$ for the bct phases derived from the group 6B elements occur in a higher density of $d$ bands relative to the $\varepsilon_F$ in the corresponding bcc phases.

### B. Plasma Frequencies and Fermi Velocities

$N(\varepsilon)<v^2>$ is shown plotted as a function of energy for all the studied metals in Figs. 3(a) – 3(f). The calculated values of $N(\varepsilon_F)<v^2>$ are shown in Table II. The highest value among the bcc elements is obtained for Ta (5.290 Ry. bohr$^2$), and the lowest value for Nb (1.572 Ry. bohr$^2$). Among the group 5B elements, $N(\varepsilon_F)<v^2>$ values for the bct phases of V and Ta increased by 4% and 66% respectively over the bcc values, while the $N(\varepsilon_F)<v^2>$ value for bct Nb decreased by 10% from the bcc value. Among the group 6B elements, $N(\varepsilon_F)<v^2>$ values for the bct phases of W, Mo and Cr decreased by 96%, 58% and 37% respectively from the bcc values. We have also calculated the square root of the mean squared Fermi velocity, $<v^2>^{1/2}$, for all the elements in their bcc and bct phases from Eq. (10) and Eq. (13), assuming a spherical Fermi surface. The highest value of $<v^2>^{1/2}$ (Table II) among the bcc elements is obtained for W (9.92 x 10$^7$ cm/s), and the lowest value for Nb (2.97 x 10$^7$ cm/s). Our calculated values of $<v^2>^{1/2}$ are compared in Table II to other values of $<v^2>^{1/2}$ obtained by the electron lifetime model and the APW band-structure method,[63] Slater-Koster interpolation of APW data,[35] and experimental measurement.[62] Our calculated $<v^2>^{1/2}$ for the group 5B elements V and Ta and the group 6B elements Mo and W differ by an average of 11% and 3% respectively from these previously reported values. However, our calculated value of $<v^2>^{1/2}$ for Nb differed by 70% from the literature reported value.[62]

In each studied period, the group 6B bcc metal always had a higher value of $<v^2>^{1/2}$ than the group 5B bcc metal in the same period. Bct values of $<v^2>^{1/2}$ increased relative to the corresponding bcc phase values for the group 5B elements, while the bct values of $<v^2>^{1/2}$ decreased relative to the corresponding bcc phase values for the group 6B elements.

The plasma frequencies calculated for all the studied metals according to Eq. (9) and Eq. (10) are also shown in Table II. The highest magnitude of $\omega$ is obtained for cubic Mo (8.71 eV), while the lowest is obtained for cubic Nb (4.48 eV). Our values are compared in Fig. 4 to values obtained experimentally and theoretically by other authors and are mostly well within the ranges reported by these authors.

### C. Electrical resistivity

Experimental resistivity data are compared to our calculated resistivities over the temperature range 20 – 700 K in Figs. 5(a) – 5(e). At room temperature, the calculated resistivities differ from the experimental data by less than 20% for all the studied bcc metals except for Nb, for which an anomalously large value of the resistivity (200% greater than experiment) is obtained from our computations. Over the temperature range 300 – 700 K, the average discrepancies between experimental and calculated resistivities are 9% (Ta), 250% (Nb), 19% (V) and 10% (Mo and W). The order of resistivity magnitudes seen in the experimental data at room temperature is reproduced by our calculations if the calculated value for Nb is not considered.

# V. DISCUSSION

Metastable phases of vanadium have been prepared experimentally and correlated using LEED to a calculated metastable bct phase produced by tetragonal distortion.[1,2] The tetragonal distortion was induced by depositing vanadium on a Ni {001} substrate which had a lattice constant 0.8% smaller than the theoretically predicted square side of tetragonal vanadium. The high density of defects and disorder in the experimentally prepared sample of metastable vanadium was attributed to the low energy difference per unit cell (0.14 eV) separating the calculated stable and metastable phases. In this work, the energy differences separating the calculated stable and metastable phases ranged from 0.09 eV (vanadium) to 0.39 eV (tungsten). It may be possible that all the metastable phases predicted in this paper by tetragonal distortion of the corresponding bcc phases can be prepared experimentally in a manner similar to that related above for vanadium. There could then exist a whole range of metastable thin film structural phases, stabilized by suitable substrates and derived from the bct phases predicted in this study by substrate lattice parameter dependent strains. The strains would also determine the mechanical and structural properties of the metastable phases. Such metastable film phases might also be expected to transform to the stable cubic phases upon heating, since the calculated energy differences between their tetragonal and cubic phases do not differ appreciably from that of vanadium, for which metal this characteristic has been observed by Tian *et al*.[1] It is not known with certainty if the substrate strain mediated process discussed above for forming metastable phases occurs at any stage of the β–tantalum formation process. We have studied this question by comparing the resistivity calculated for our predicted bct Ta phase to values obtained from resistivity measurements on pure β–tantalum films of varying thicknesses.

**A. Densities of states**

The bcc metals in this study span periods 4, 5 and 6, and groups 5B – 6B of the Periodic Table. The calculated total densities of states, $N(\varepsilon_F)$, for the bcc metals followed a decreasing trend when proceeding down the periods and moving rightwards across the groups of the studied block of the Periodic Table. The magnitude of $N(\varepsilon_F)$ is inversely proportional to the $d$ bandwidth and directly proportional to the $d$ band energy. It also follows a trend inversely proportional to the magnitude of the atomic number $Z$. The order of the bcc $N(\varepsilon_F)$ values down the periods is due to the $d$ bandwidth, which increases in this direction. The order of the bcc $N(\varepsilon_F)$ values across the groups is due to the decrease in the $d$ band energy, due to band filling, across the groups.

The dominant effect of the open shell $d$ occupancy is expected because from Table II, the $d$ densities of states were responsible for the largest contribution to $N(\varepsilon_F)$ for all the studied bcc metals. The calculated $N(\varepsilon_F)$ for the bct phases derived from the group 5B bcc elements were consistently lower than those calculated for their corresponding bcc phases. The reverse relationship to the bcc phase $N(\varepsilon_F)$ values was seen for the $N(\varepsilon_F)$ calculated for bct phases derived from the group 6B bcc elements.

Within each studied group, the magnitude of the $c/a$ ratio at which a bct phase is formed from the bcc phase is inversely proportional to $Z$, while the size of $\Delta E^{TOT}$ is directly proportional to $Z$. Within each period, the group 6B elements formed bct phases with lower $c/a$ ratios than the group 5B elements, and had higher values of $\Delta E^{TOT}$. This dependence of the $c/a$ ratio on $Z$ and the number of outer shell electrons is exactly the same noted earlier for $N(\varepsilon_F)$, while the dependence of $\Delta E^{TOT}$ on these quantities is the exact opposite. We explain this by noting that the percent change in the fraction of bonding states at $\varepsilon_F$ for the bcc to bct phase transition is much greater (215 – 300%) for the group 5B elements than for the group 5B elements (65 – 85%). We can therefore conclude that within each studied group and across each studied period, the magnitude of the bcc $d$ bandwidth is inversely proportional to the $c/a$ ratio at which a bct phase is formed from the bcc phase and inversely proportional to $\Delta E^{TOT}$. If the magnitude of $\Delta E^{TOT}$ is taken as a measure of the degree of difficulty in forming a bct phase from the bcc structure, this suggests that heavier elements are less likely to form bct phases and that the bct phases derived from the group 6B elements are less readily formed from the corresponding bcc phases than bct phases derived from group 5B elements, because $\Delta E^{TOT}$ increases appreciably across groups.

**B. Plasma frequency**

The calculated plasma frequencies are compared to plasma frequencies obtained experimentally and theoretically by other authors in Fig. 4. Our values are well within the range obtained by these other works. We have also calculated the plasma frequencies of Nb and Ta as a function of pressure (Tables IV and V). Our results for the pressure dependence of the plasma frequency, similar to those of Neve *et al*,[33] show a slight dependence of the plasma frequency on the pressure, but we do not observe the monotonic increase of the plasma frequency with pressure as seen in that reference, however. The fact that our calculated plasma frequencies do not increase monotonically with pressure might be due to our evaluation of Eq. (10) without taking into account spin-polarization effects. By not including spin-polarization, we have averaged over the spin-up and spin-down states. The large discrepancy between our calculated Nb $<v^2>^{1/2}$ value and the Nb $<v^2>^{1/2}$ value reported in the literature[62] is also attributed to the manner in which we evaluated Eq. (10).

## C. Electrical resistivity

The electrical resistivities for all the studied metals have been calculated over the temperature range 20 – 700 K according to Eq. (4) and are shown in Figs. 5(a) – 5(e). The calculated resistivity curves appear linear over the entire temperature range and lie above the experimental curves at the lower end of the temperature scale for all the metals. The reasons for this are discussed below. Over the temperature range 300 – 700 K, the average discrepancies between experimental and calculated resistivities are 9% (Ta), 250% (Nb), 19% (V) and 10% (Mo and W). The discrepancies between the calculated resistivities and the resistivity data are attributed to the shortcomings of the LOVA model, the approximation which we have used to calculate the resistivities represented by Eq. (4), and also to the approximation of $\lambda_{tr}$ by $\lambda$. LOVA assumes that the Fermi surface displaces rigidly, and also that the electronic distribution function, and hence the electronic relaxation time, is energy independent and isotropic, the angular variation being proportional to $v_{kx}$.[54] Higher-order corrections to LOVA aim to incorporate the angular and energy dependence of the distribution function, as well as the effects of allowing different sheets of the Fermi surface to displace differently, into resistivity calculations. Resistivity curves calculated using LOVA usually lie above the experimental curves at temperatures below 100K, and refinements to LOVA such as the n-sheet approximation, which allow different sheets of the Fermi surface to have different velocities, lower the calculated resistivity relative to experiment in the low temperature region.[37] The calculated resistivities appear linear because the linear term in Eq. (4) dominates the other terms over the entire temperature range. The approximation of $\lambda_{tr}$ by $\lambda$ neglects the factor $\left[1-\frac{v_{kx} \cdot v_{k'x}}{|v_{kx}|^2}\right]$, which preferentially weights backscattering processes. The transition metals possess very complex, highly nested Fermi surfaces; therefore backscattering contributions to $\lambda_{tr}$ could be significant and lead to $\lambda_{tr}$ being significantly different from $\lambda$.[58]

The calculated temperature dependent resistivity for Nb had the worst agreement with experiment of all the metals over the entire range of temperatures. We attribute this to the following factors in addition to those mentioned above: 1. Our calculated value of $N(\varepsilon_F)<v^2>$ for Nb, being smaller than usual,[33] causes the calculated resistivity curve to shift significantly upward from the experimental curve. The large error in the value of $N(\varepsilon_F)<v^2>$ calculated for Nb is probably due to the fact, previously noted in this paper, that Eq. (10) was evaluated without taking into account spin-polarization effects. 2. Superconductive tunneling experiments on Nb single crystals have revealed that the Nb spectral function $\alpha^2(\omega)F(\omega)$ is strongly anisotropic.[60] Due to this anisotropy, the errors introduced into the resistivity calculation by using LOVA could be more severe for Nb than for the rest of the studied metals. When the anisotropy and energy dependence of the distribution function are accounted for in the Nb resistivity calculation, the calculated resistivities below 100 K are appreciably reduced relative to experiment and the LOVA results, while the calculated resistivities at higher temperatures are negligibly affected.[37]

3. The measured temperature dependent resistivity of Nb is known to exhibit a significant negative deviation from linearity at high temperatures.[55] LOVA and higher order corrections to it neglect Fermi smearing effects, which are significant at higher temperatures, and which along with anharmonicity, are responsible for considerable negative deviations from linearity in the resistivity.[37]

The metals with the largest values of the calculated room temperature resistivity are Nb and V, while the lowest values are obtained for W and Mo. The experimental resistivity data mirror this trend (Figs. 5). Within each group the order of the resistivity magnitudes is as follows $\rho_{3d} > \rho_{4d} > \rho_{5d}$. The in-group order of resistivities is due to the magnitude of the $d$ conduction bandwidth, which increases down the periods. $N(\varepsilon_F)$ is largely of $d$ character for all the studied metals, so an increase in $d$ bandwidth should correspond to a greater average velocity of electrons on the Fermi surface. This is confirmed by our calculated values of $<v^2>^{1/2}$, which except for the Nb value, also increase in this direction. The Fermi energies of the group 5B metals V, Nb and Ta occur inside a high density of $d$ bands, unlike the Fermi energies of the group 6B metals Cr, Mo and W, which occur near a minimum in the DOS (Figs. 2). This means that the states in the range of the Fermi energy are more diffuse in the group 6B metals and have little contribution inside the muffin-tin spheres. The greater extent of the electrons with energies near $\varepsilon_F$ in the group 6B bcc elements is again confirmed by our calculated values of $<v^2>^{1/2}$, which as was previously noted, are higher in each period for the group 6B elements than for the group 5B elements. The room temperature resistivities of the group 6B metals are all less than those of the group 5B metals, so the combined effects of the $d$ bandwidth and the position of $\varepsilon_F$ in the band structure determine the order of resistivities. Applying this logic, the bct phases derived from the group 5B metals, including bct Ta, should in general be more conductive than their corresponding bcc phases, since the $\varepsilon_F$ positions in the bct phases are shifted to lower DOS, while the $d$ bandwidths are approximately the same as the bcc $d$ bandwidths. Conversely the bct phases derived from the group 6B metals should generally be less conductive than their corresponding bcc phases, since the $\varepsilon_F$ positions in these phases are shifted to higher DOS, without any appreciable changes in the bct $d$ bandwidths relative to those of the bcc phases. These general conclusions can be confirmed for some of the metals by comparing the values of $N(\varepsilon_F)<v^2>$ and $<v^2>^{1/2}$ calculated for the bcc and bct phases. As noted in section III, among the group 6B elements, $N(\varepsilon_F)<v^2>$ values for the bct phases of W, Mo and Cr decreased by 96%, 58% and 37% respectively from the bcc values. The resistivities of the bct phases predicted for the group 6B metals Cr, Mo and W should therefore increase appreciably over the respective bcc phase values not only according to Eq. (1), but also because the calculated bct value of $<v^2>^{1/2}$ decreases from the bcc value by 39%, 48% and 86% for Cr, Mo and W respectively. Among the group 5B elements, $N(\varepsilon_F)<v^2>$ values for the bct phases of V and Ta increased by 4% and 66% respectively over the bcc values, while the $N(\varepsilon_F)<v^2>$ value for bct Nb decreased by 10% from the bcc value. The calculated bct value of $<v^2>^{1/2}$ increases over the bcc value by 17%, 300% and 32% for V, Nb and Ta respectively. The bct phases predicted for the studied group 5B metals V, Nb and Ta should therefore be more conductive than the respective bcc phases.

The calculated resistivities (except in the case of Nb) differ from experimental data by at most 20% and follow the trend in the order of magnitudes seen in the experimental data. The measured β–Ta resistivity is 1000 - 2000 % greater than that of α–Ta, therefore we believe that our calculations can determine if the frequently observed β phase of tantalum has the bct structure predicted in this paper.

There is some evidence that impurities such as oxygen may stabilize the beta phase of tantalum.[7,25,26,29-31] Metastable phases of tantalum have been observed to readily transform to the stable cubic phase upon heating, with a significant decrease in resistivity and oxygen impurity content.[18,27] It has therefore been suggested[57] that the relatively high resistivity of beta tantalum is due to the effects of incorporated impurities and its small grain size, which was found to be about 45% less than that of bcc tantalum. However studies of ion-facilitated metal film growth[59] have concluded that the decrease in grain size from bcc to beta tantalum should only account for a twofold increase in resistivity, not the nearly tenfold increase that is observed. Furthermore, metastable tantalum phases which have been prepared with low levels of contamination still exhibit high values of resistivity,[24,25] therefore the experimentally measured high resistivity of beta tantalum is believed to be due less to impurities and grain size effects, and more to the intrinsic transport properties of its crystalline structure. Thermally induced structural transformation to the stable bcc phase has also been observed with metastable phases of tungsten.[4] There have however been conflicting reports regarding the oxygen content of bcc tungsten formed by heating relative to its metastable precursor,[4,6] which is known to be about three times more resistive than bcc tungsten.[20,21]

In order to determine whether the bct phase of Ta predicted in this work is produced at any stage of the β–Ta formation process, we compare the calculated resistivity of the bct phase with resistivity values obtained from measurements on pure β–Ta films of varying thicknesses. We can carry out the comparison in two ways. The calculated transport properties of bct Ta and the measured resistivities of β–Ta can be used to obtain a range of empirical values for $\lambda_{tr}$, which should be on the order of unity. Alternatively, reasonable values of $\lambda_{tr}$ for β–Ta can be used in Eq. (3), together with the computed transport properties of bct Ta to directly compute the resistivity, which can be compared to the resistivity values measured for β–Ta. Using the value of $N(\varepsilon_F)<v^2>$ computed for bct Ta and the measured resistivity of β–Ta at 400 K[45] in Eq. (3), we obtain an empirical $\lambda_{tr}$ value of 18.7 for β–Ta, which is clearly unreasonable. Conversely, using a reasonable range of $\lambda_{tr}$ values (0.5 – 1) in Eq. (3) together with the value of $N(\varepsilon_F)<v^2>$ computed for bct Ta results in a calculated bct Ta resistivity range of 7.16 – 14.34 μΩ cm, which is less than the measured value (18.22 μΩ cm) and the calculated value (19.78 μΩ cm) for bcc Ta. This does not agree with numerous experimental findings discussed previously. It is therefore highly probable that the bct tantalum phase predicted in this work is not the same metastable (β) tantalum phase that has been observed in so many experiments.

# VI. CONCLUSION

The bcc metals in groups 5B and 6B, spanning periods 4 - 6 have been studied using the FP-LAPW band structure method. Within each studied group and across each studied period, the magnitude of the bcc $d$ bandwidth is inversely proportional to the $c/a$ ratio at which a bct phase is formed from the bcc phase and inversely proportional to $\Delta E^{TOT}$, the energy difference between corresponding bcc and bct phases. Heavier elements are less likely to form bct phases and the bct phases derived from the group 6B elements are less readily formed from the corresponding bcc phases than bct phases derived from group 5B elements. Bct phases formed from group 5B elements are more conductive than the bcc phases, unlike bct phases formed from group 6B elements.

A metastable bct phase of tantalum with a $c/a$ ratio of 1.7 has been predicted. This bct phase is not produced at any stage of the β–tantalum formation process. The calculated resistivities of this bct phase and the bcc tantalum phase show that the observed high resistivity of β–tantalum relative to the bcc phase, which is impurity independent, is not due to simple tetragonal distortions of the bcc phase, such as were used in this paper to produce a metastable bct tantalum phase. The relatively high resistivity observed experimentally for β–tantalum is likely due to the intrinsic transport properties, which are largely determined by the Fermi surface, of a crystalline structure more complex than that formed by a simple tetragonal distortion of bcc tantalum. Other proposed structures[7,8,24] of β–tantalum are currently being studied to further clarify the details of the Fermi surface effect on the resistivity.

**ACKNOWLEDGEMENTS**

The authors gratefully acknowledge support from the U. S. Army Sustainable Green Manufacturing Program. NON gratefully acknowledges support from a National Science Foundation MAGNET fellowship.

**FIGURE CAPTIONS**

**FIGS. 1(a) - 1(f)** The total energy per unit cell as a function of $c/a$ ratio for the group 5B elements Ta, Nb and V, and the group 6B elements W, Mo and Cr. The total energies shown for tantalum, niobium and vanadium are with respect to $-31252$ Ry, $-7634$ Ry and $-1898$ Ry respectively. The total energies shown for tungsten, molybdenum and chromium are with respect to $-32332$ Ry, $-8099$ Ry and $-2101$ Ry respectively.

**FIGS. 2(a) - 2(f)** The densities of states as functions of energy are shown for the cubic and the tetragonal phases of the group 5B elements Ta, Nb and V, and the group 6B elements W, Mo and Cr. In each of Figs. 2(a) to 2(f), the Fermi energy is at 0 eV. In the upper panel of each figure, the thickened full lines and dashed lines represent respectively the total and $d$ densities of states for the tetragonal phase. The thin full and dashed lines represent respectively the total and $d$ densities of states for the cubic phase. In the lower panel of each figure, the thickened full and dashed lines represent respectively, the $d_{z^2}$ and the $d_{xz+yz}$ densities of states. The thin full and dashed lines represent respectively the $d\text{-}e_g$ and the $d\text{-}t_{2g}$ densities of states.

**FIGS. 3(a) –3(f)** $N(E)<v^2>$ as a function of energy is shown for all the studied group 5B and group 6B metals.

**FIG. 4.** Plasma frequency values obtained experimentally and theoretically by various authors for the same metals in this study are plotted and compared with the plasma frequency values calculated in this work.

**FIGS. 5(a) – 5(e)** Measured and calculated values of the electrical resistivity over the temperature range 20 –700 K are shown for the group 5B metals Ta, Nb and V, and the group 6B metals Mo and W.

**TABLE I.** The calculated $c/a$ ratios of the predicted metastable phases of the cubic metals studied, the lattice parameter and the calculated total energy per unit cell for both phases of each metal, and the calculated energy differences $\Delta E^{TOT}$ between the phases for each metal are tabulated.

| Metal | $a_{bcc}$ (Å) | $a_{bct}$ (Å) | $c/a$ ratio of bct phase | $E_{bcc}^{TOT}$ (Ry) | $E_{bct}^{TOT}$ (Ry) | $\Delta E^{TOT}$ (eV) |
|---|---|---|---|---|---|---|
| Ta | 3.301 | 2.765 | 1.70 | -31252.2369 | -31252.225 | 0.16 |
| Nb | 3.300 | 2.723 | 1.78 | -7640.9526 | -7640.9421 | 0.14 |
| V  | 3.030 | 2.491 | 1.80 | -1898.6472 | -1898.6407 | 0.09 |
| W  | 3.165 | 2.662 | 1.68 | -32332.2763 | -32332.2476 | 0.39 |
| Mo | 3.147 | 2.611 | 1.75 | -8099.1836 | -8099.1618 | 0.30 |
| Cr | 2.910 | 2.400 | 1.78 | -2101.7819 | -2101.7640 | 0.24 |

**TABLE II.** The calculated total density of states at the Fermi energy, $N(\varepsilon_F)$, the total $d$ density of states at the Fermi energy, $N_d(E_F)$, the product of the total density of states at the Fermi level and the mean square Fermi velocity, $N(E_F)<v^2>$, as well as the square root of the mean square Fermi velocity, $<v^2>^{1/2}$, are shown for bct and bcc phases of all the studied metals. The numbers in brackets next to the bcc $<v^2>^{1/2}$ values are values of $<v^2>^{1/2}$ obtained by other authors. The calculated plasma frequency values, $\hbar\omega$, are also shown for all the bcc metals. See Fig. 4 for comparison of the $\hbar\omega$ calculated in this paper with previous work.

| Metal | c/a ratio | $N(E_F)$ (States/eV.atom) | $N_d(E_F)$ (States/eV.atom) | $N(E_F)<v^2>$ (Ry.bohr$^2$) | $<v^2>^{1/2}$ ($10^7$ cm/s) | $\hbar\omega$ (eV) |
|---|---|---|---|---|---|---|
| Ta | 1.00 | 1.31 | 0.81 | 5.290 | 5.96 (6.70)[a,t] | 8.22 |
|    | 1.70 | 1.25 | 0.74 | 8.801 | 7.87 | |
| Nb | 1.00 | 1.57 | 0.93 | 1.572 | 2.97 (5.10)[b,e] | 4.48 |
|    | 1.78 | 1.18 | 0.70 | 1.417 | 12.00 | |
| V  | 1.00 | 1.99 | 1.42 | 2.619 | 3.40 (3.73)[c,t] | 6.58 |
|    | 1.80 | 1.51 | 0.99 | 2.727 | 3.99 | |
| W  | 1.00 | 0.38 | 0.25 | 4.249 | 9.92 (9.6)[a,t] | 7.85 |
|    | 1.68 | 0.72 | 0.51 | 0.161 | 1.40 | |
| Mo | 1.00 | 0.58 | 0.41 | 5.150 | 8.84 (8.6)[a,t] | 8.71 |
|    | 1.75 | 0.91 | 0.66 | 2.186 | 4.60 | |
| Cr | 1.00 | 0.68 | 0.56 | 2.208 | 5.34 | 6.42 |
|    | 1.80 | 1.16 | 0.87 | 1.382 | 3.24 | |

[a]Ref. 35.
[b]Ref. 62.
[c]Ref. 63.
e – experimental, t – theoretical.

**TABLE III.** The total $d$ densities of states at the Fermi energy, $N_d(E_F)$, as well as the symmetry decomposed components of $N_d(E_F)$ are shown for the bcc and bct phases of all the studied metals.

| Metal | $N_l(E_F)$ | | | | | | | |
|---|---|---|---|---|---|---|---|---|
| | $d_{bcc}^{tot}$ | $d_{e_g}$ | $d_{t_{2g}}$ | $d_{bct}^{tot}$ | $d_{z^2}$ | $d_{x^2-y^2}$ | $d_{xy}$ | $d_{xz}+d_{yz}$ |
| Ta | 0.81 | 0.14 | 0.67 | 0.74 | 0.20 | 0.30 | 0.06 | 0.18 |
| Nb | 0.93 | 0.18 | 0.75 | 0.70 | 0.16 | 0.26 | 0.07 | 0.21 |
| V  | 1.42 | 0.24 | 1.18 | 0.99 | 0.26 | 0.39 | 0.08 | 0.26 |
| W  | 0.25 | 0.05 | 0.20 | 0.51 | 0.12 | 0.07 | 0.09 | 0.23 |
| Mo | 0.41 | 0.11 | 0.30 | 0.66 | 0.12 | 0.10 | 0.13 | 0.31 |
| Cr | 0.53 | 0.12 | 0.41 | 0.88 | 0.16 | 0.17 | 0.13 | 0.42 |

**TABLE IV.** The pressure dependence of $N(E_F)<v^2>$, the plasma frequency $\hbar\omega$, the Fermi energy $E_F$, and the density of states at the Fermi energy $N(E_F)$ are shown for bcc niobium.

| $\dfrac{V}{V_0}$ | $a_{Nb}$ (bohr) | $N(E_F)<v^2>$ (Ry.bohr$^2$) | $\hbar\omega$ (eV) | $E_F$ (Ry) | $N(E_F)$ (states/Ry.atom) |
|---|---|---|---|---|---|
| 1.04 | 6.319 | 1.815 | 4.72 | 0.733 | 22.48 |
| 1.03 | 6.299 | 1.783 | 4.70 | 0.743 | 22.08 |
| 1.02 | 6.278 | 1.741 | 4.67 | 0.752 | 21.81 |
| 1.01 | 6.258 | 1.584 | 4.48 | 0.762 | 21.55 |
| 1.00 | 6.237 | 1.572 | 4.48 | 0.807 | 21.14 |
| 0.99 | 6.216 | 1.574 | 4.50 | 0.818 | 20.88 |
| 0.98 | 6.195 | 1.503 | 4.43 | 0.829 | 20.48 |
| 0.97 | 6.174 | 1.391 | 4.28 | 0.839 | 20.07 |
| 0.95 | 6.131 | 1.342 | 4.25 | 0.862 | 19.54 |

**TABLE V.** The pressure dependence of $N(E_F)<v^2>$, the plasma frequency $\hbar\omega$, the Fermi energy $E_F$ and the density of states at the Fermi energy $N(E_F)$ are shown for bcc tantalum.

| $\dfrac{V}{V_0}$ | $a_{Ta}$ (bohr) | $N(E_F)<v^2>$ (Ry.bohr$^2$) | $\hbar\omega$ (eV) | $E_F$ (Ry) | $N(E_F)$ (states/Ry.atom) |
|---|---|---|---|---|---|
| 1.04 | 6.321 | 5.323 | 8.09 | 0.760 | 18.87 |
| 1.03 | 6.300 | 5.183 | 8.02 | 0.769 | 18.74 |
| 1.02 | 6.280 | 5.138 | 8.02 | 0.779 | 18.60 |
| 1.01 | 6.259 | 5.060 | 8.00 | 0.789 | 18.33 |
| 1.00 | 6.239 | 4.948 | 7.95 | 0.798 | 18.11 |
| 0.99 | 6.218 | 4.710 | 7.80 | 0.825 | 17.93 |
| 0.98 | 6.197 | 4.670 | 7.80 | 0.835 | 17.80 |
| 0.97 | 6.176 | 4.691 | 7.86 | 0.846 | 17.53 |
| 0.95 | 6.133 | 4.576 | 7.84 | 0.867 | 17.01 |

**Fig. 1 (a)**

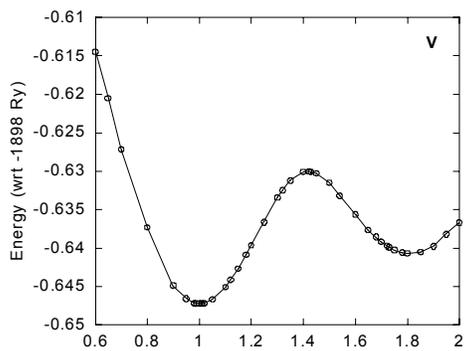

**Fig. 1 (b)**

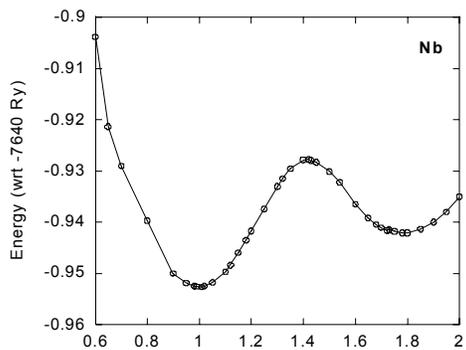

**Fig. 1 (c)**

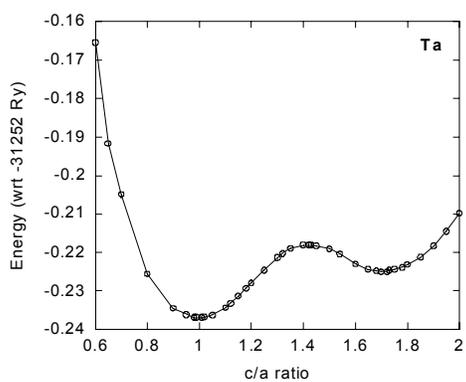

**Fig. 1 (d)**

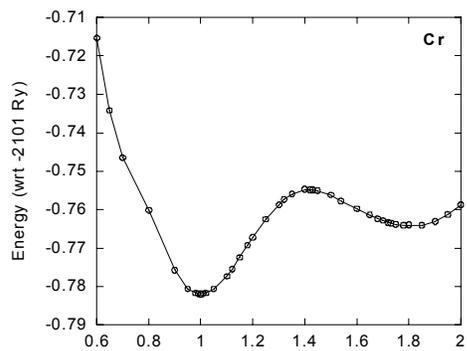

**Fig. 1 (e)**

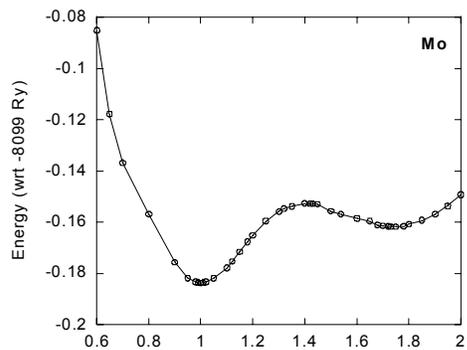

**Fig. 1 (f)**

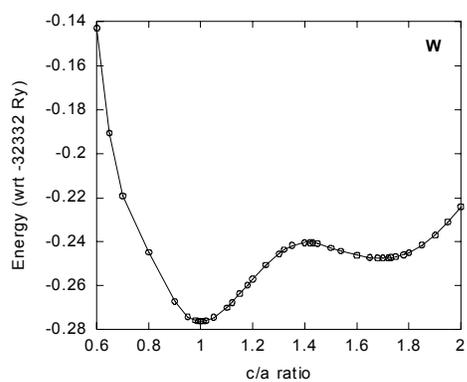

**Fig. 2 (a)**

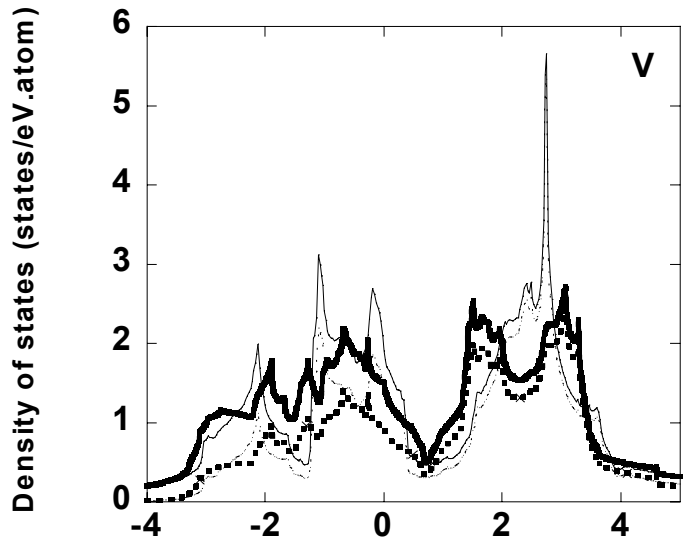

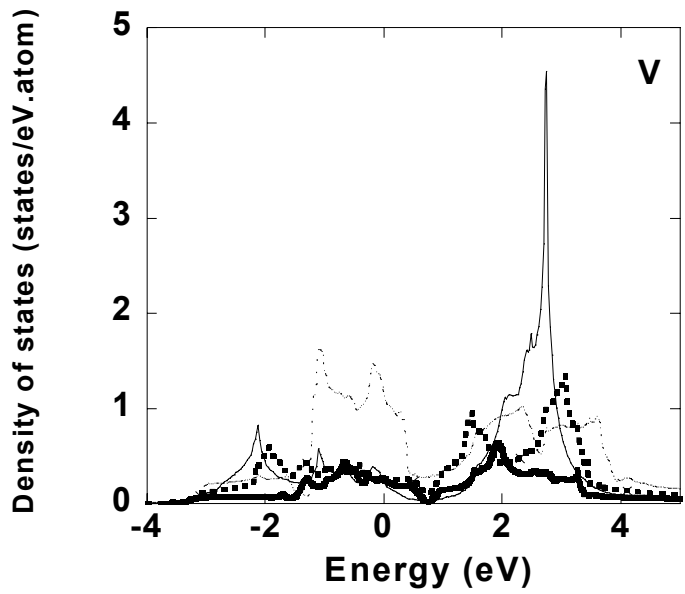

**Fig. 2 (b)**

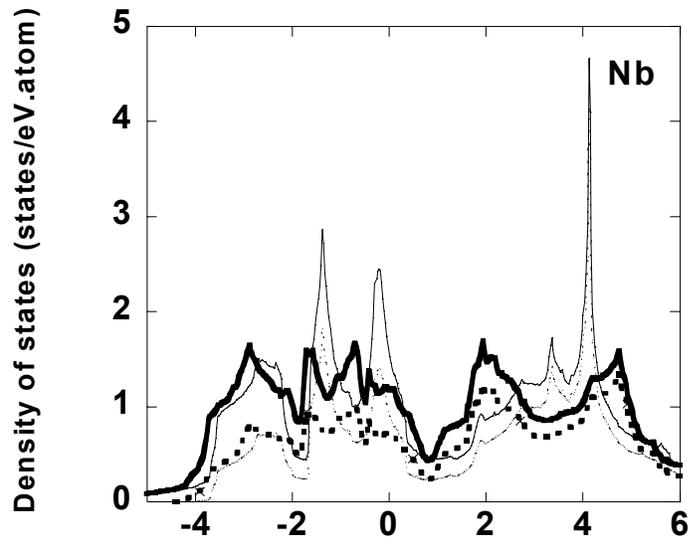

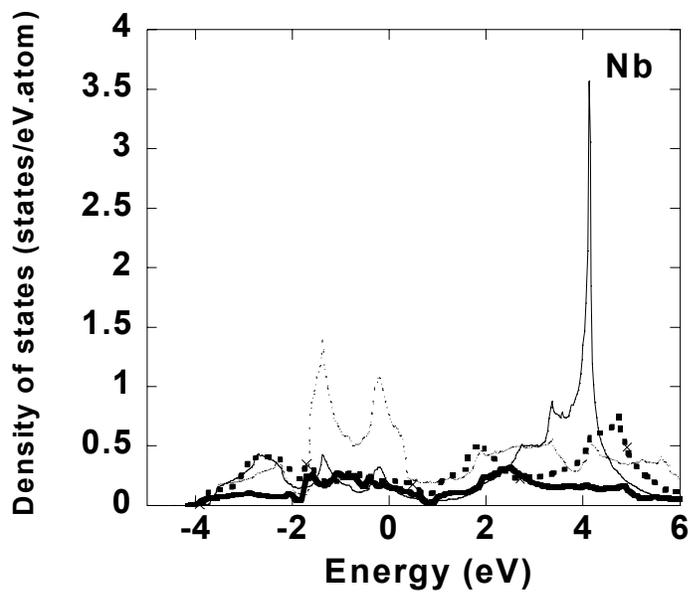

**Fig. 2 (c)**

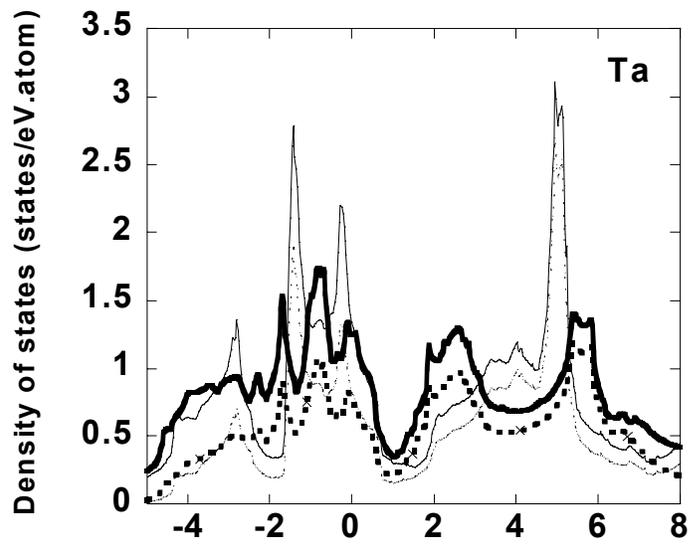

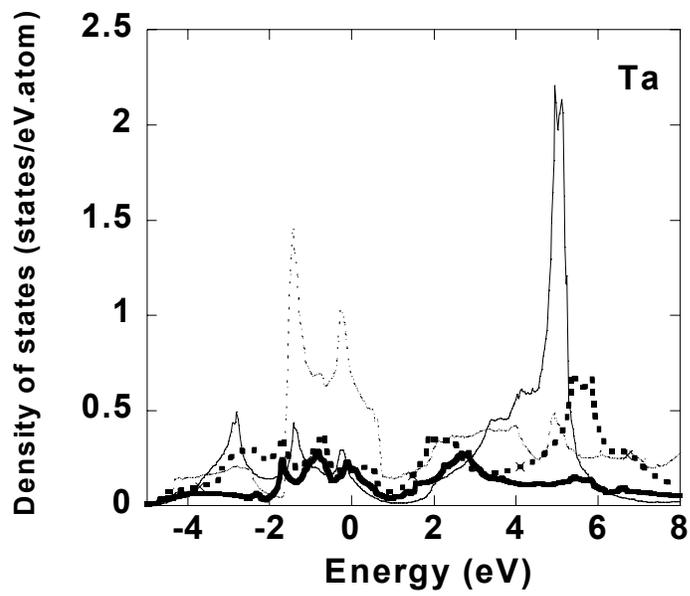

**Fig. 2 (d)**

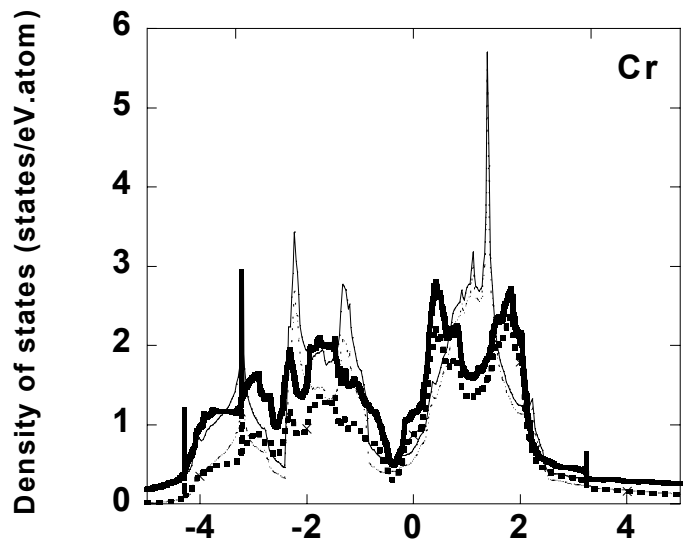

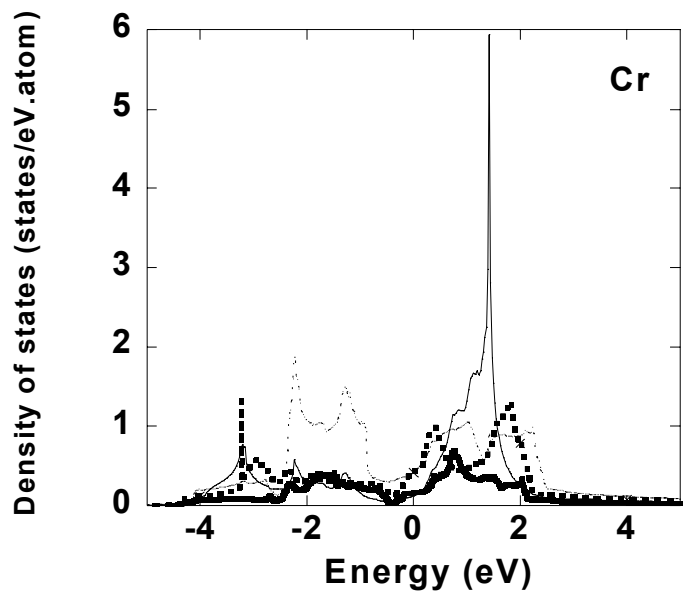

Fig. 2 (e)

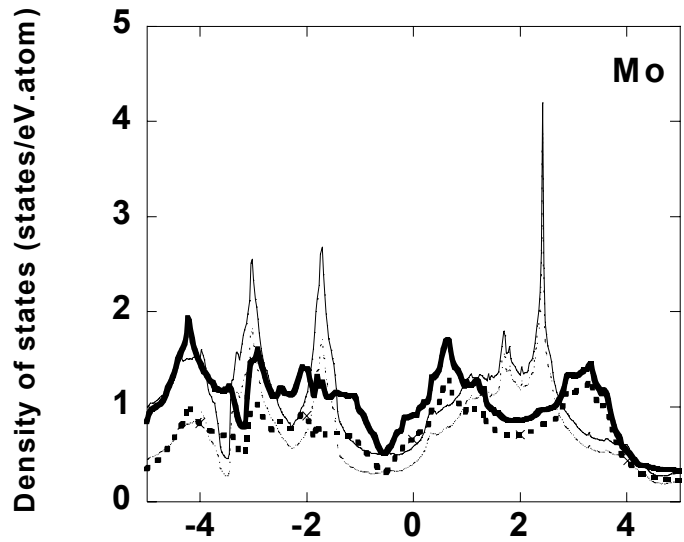

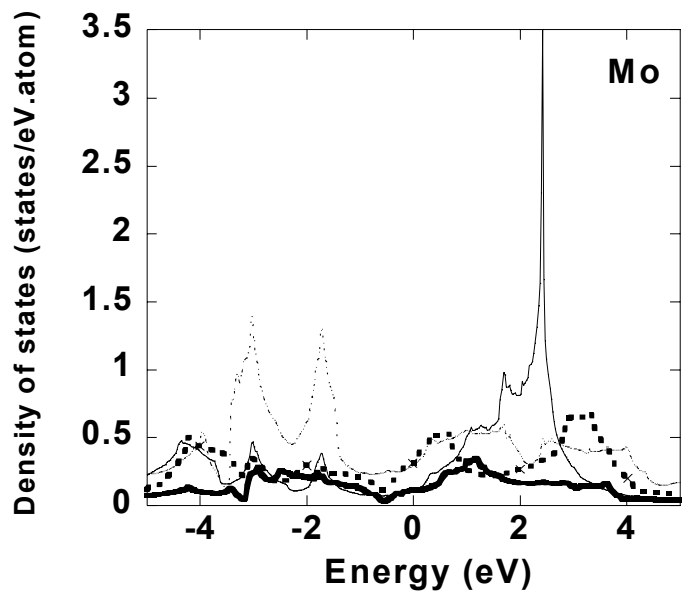

Energy (eV)

**Fig. 2 (f)**

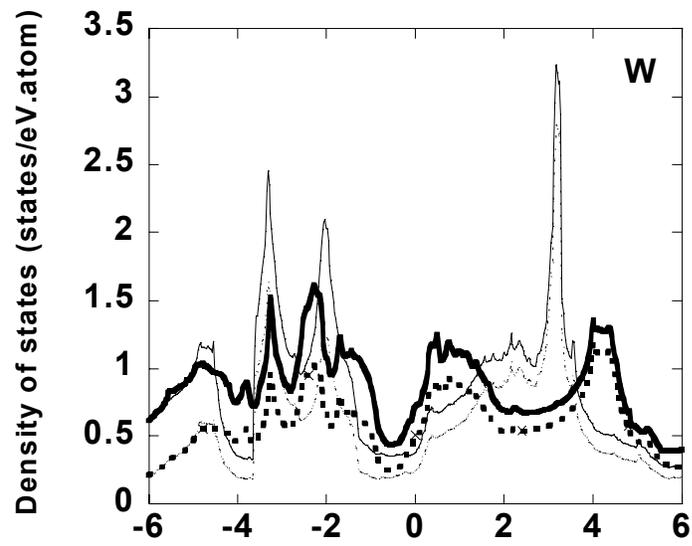

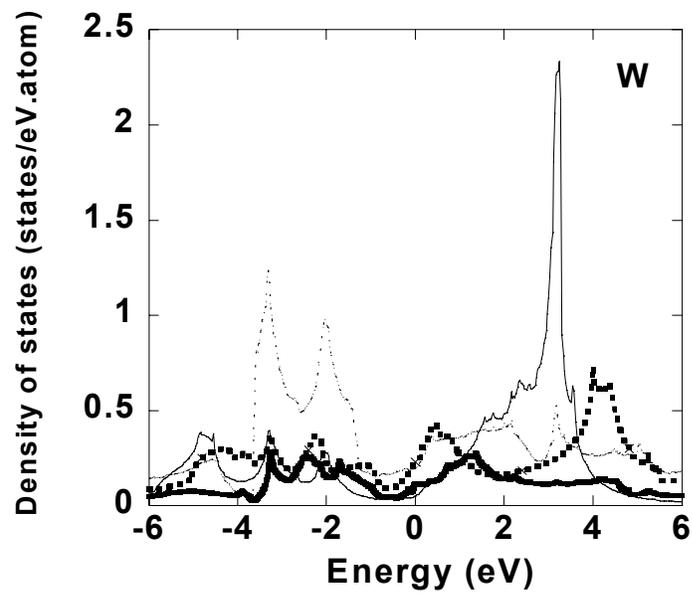

**Fig. 3 (a)**

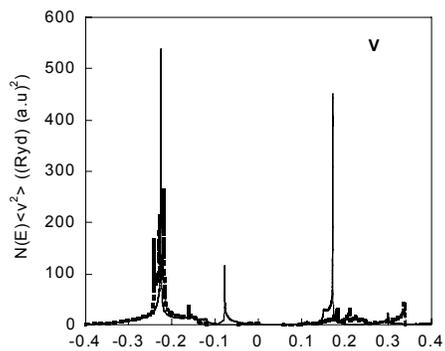

**Fig. 3 (b)**

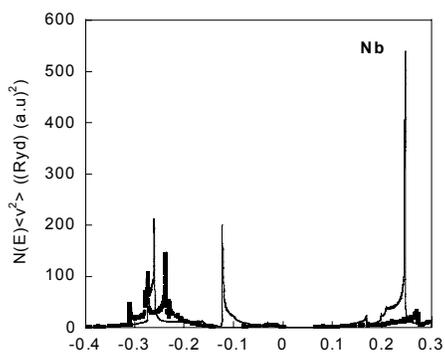

**Fig. 3 (c)**

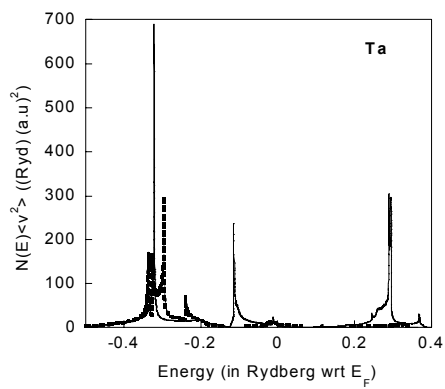

**Fig. 3 (d)**

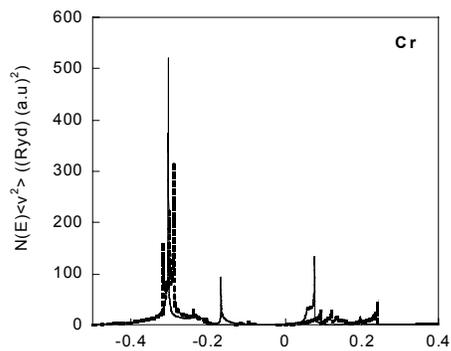

**Fig. 3 (e)**

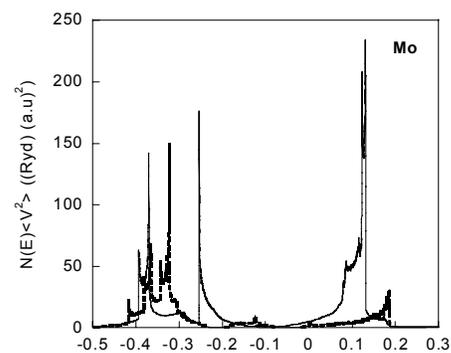

**Fig. 3 (f)**

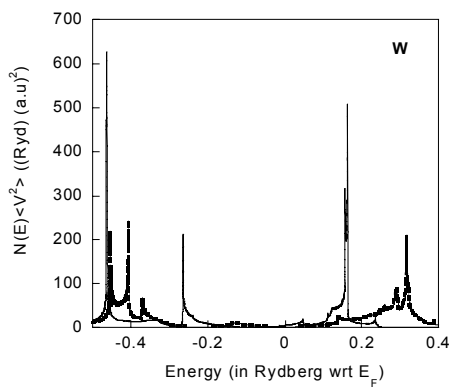

**Fig. 4**

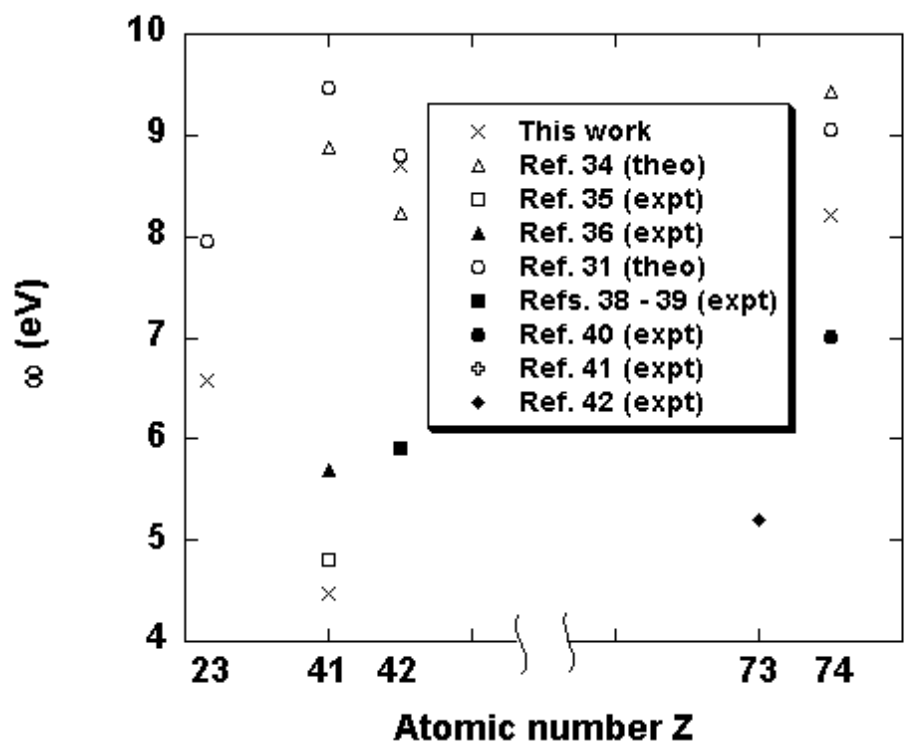

**Fig. 5 (a)**

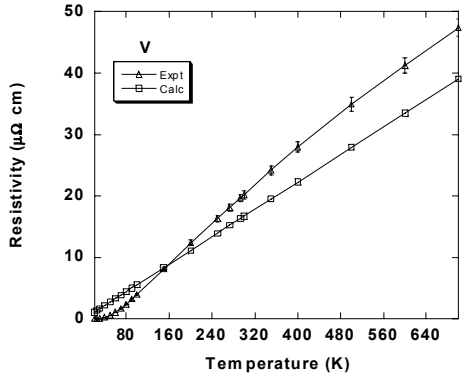

**Fig. 5 (b)**

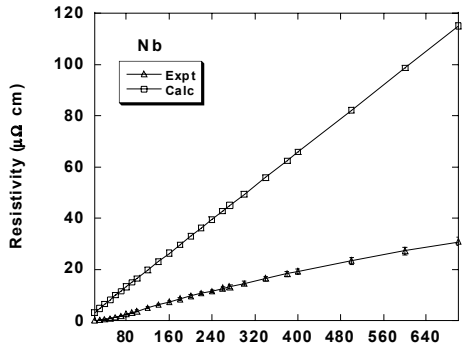

**Fig. 5 (c)**

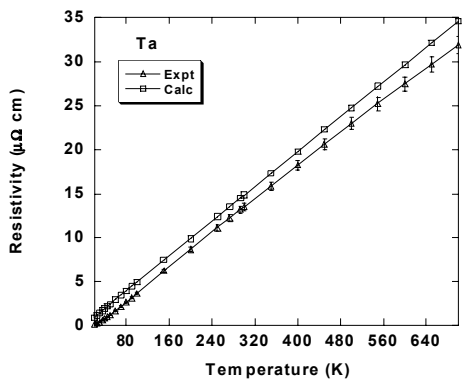

**Fig. 5 (d)**

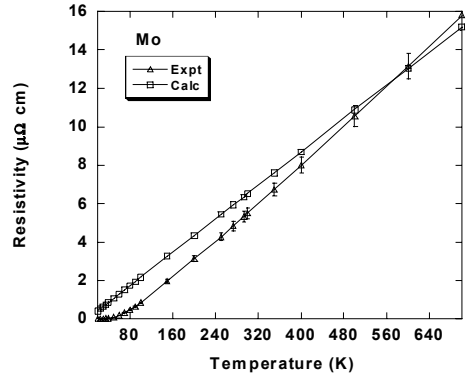

**Fig. 5 (e)**

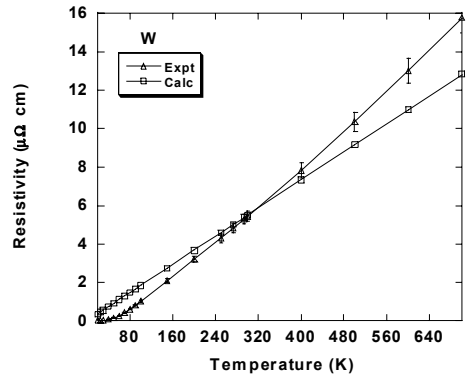